\documentclass[aps,prb, twocolumn,superscriptaddress]{revtex4-2}

\usepackage[utf8]{inputenc}
\usepackage{amsthm}
\usepackage{amsmath}
\usepackage{xfrac}
\usepackage{color}
\usepackage{soul,xcolor}
\usepackage{hyperref}
\usepackage{graphicx,xcolor}
\usepackage{amssymb}
\usepackage{bbold}
\usepackage{gensymb}
\usepackage{dsfont}
\usepackage{float}

\setstcolor{red}

\usepackage[normalem]{ulem}
\newcommand{\commentout}[1]{}

\newcommand{\ii}{\mathrm{i}} 
\newcommand{\diff}{\mathrm{d}} 
\newcommand{\id}{\mathbb{1}} 

\newcommand{\ket}[1]{|#1\rangle} 
\newcommand{\bra}[1]{\langle#1|} 
\newcommand{\braket}[1]{\langle#1\rangle} 
\newcommand{\ketbra}[2]{|#1\rangle\!\langle #2|} 
\newcommand{\tr}{\mathrm{tr}} 

\begin{document}

\title{Exact Floquet dynamics of strongly damped driven quantum systems}

\author{Konrad Mickiewicz}
\affiliation{Institut f{\"u}r Theoretische Physik, Technische Universit{\"a}t Dresden, 
D-01062, Dresden, Germany}

\author{Valentin Link}
\affiliation{Institut für Physik und Astronomie, Technische Universität Berlin, D-10623, Berlin, Germany}

\author{Walter T. Strunz}
\affiliation{Institut f{\"u}r Theoretische Physik, Technische Universit{\"a}t Dresden, 
D-01062, Dresden, Germany}

\begin{abstract}
We present an approach for efficiently simulating strongly damped quantum systems subjected to periodic driving, employing a periodic matrix product operator representation of the influence functional. This representation enables the construction of a numerically exact Floquet propagator that captures the non-Markovian open system dynamics, thus providing a dissipative analogue to the Floquet Hamiltonian of driven isolated quantum systems. We apply this method to study the asymptotic heating of a reservoir in spin-boson models, characterizing the deviation from equilibrium conditions. Moreover, we show how a local driving of two qubits can be utilized to stabilize a transient entanglement buildup of the qubits originating from the interaction with a common environment. Our results make it possible to directly study both stationary and transient dynamics of strongly damped and driven quantum systems within a transparent theoretical and numerical framework.
\end{abstract}\maketitle

\paragraph*{Introduction}

When a closed physical system is subjected to an external driving, the energy of the system will typically increase, as the work performed by, for instance, a drive laser is absorbed \cite{ikeda2021Fermis}. This heating process poses a general obstacle for Floquet engineering, that is engineering desired states or dynamics through periodic driving \cite{eckardt2017Colloquium, reitter2017Interaction,chen2025Mitigating}. Stable long-time evolution can be realized if the system is able to dissipate the energy from the drive into its environment. In this situation a quasi-stationary nonequilibrium state is stabilized by the interplay of drive and dissipation \cite{Shirai2016May,engelhardt2019Discontinuities,Sato2020Oct, petiziolCavityBasedReservoirEngineering2022a, Mori2023Mar, li2024Coherent, Ritter2025}. For quantum systems a theoretical description of the dissipation alone is a challenging task, as quantum correlations between system and environment need to be considered when the dissipation becomes strong or the environment has a nontrivial structure \cite{deVega2017Jan,Hartmann2020,tello2024Benchmarking}. The problem is further complicated if a strong driving is applied, as textbook master-equation approaches need to be refined and the typical simplifying approximations loose their validity \cite{Mori2023Mar,gulacsi2025Temporally}. However, a reliable and broadly accessible theoretical description of strongly driven and damped quantum systems is highly desired \cite{Kohler1997Jan, Marthaler2006, Ketzmerick2010Aug, Lazarides2020Apr, Gunderson2021Feb, Shen2022Feb, Boness2024, direkci2025Universality} due to the relevance of these systems in quantum engineering and quantum thermodynamics \cite{koyanagi2022Numerically,boettcher2024Dynamics,kolisnyk2024Floquet,Alamo2024Jun}, as well as the current perspectives for experimental realizations \cite{Koski2018Jul,vonHorstig2024Jul,Frattini2024,Ritter2025,ivakhnenko2025Probing}. 
In this work we introduce a framework that can be used to simulate exact dynamics of driven open quantum systems via time-periodic representations of influence functionals (Floquet-IF). Our method builds upon the time-translation invariant semi-group form of influence functionals~\cite{Link2024May,sonner2025Semi} which we generalize to periodically driven systems. We demonstrate the utility of this approach through applications to driven spin-boson models, where we quantify energy transfer from the driving field to the reservoir and show how driving can be used to stabilize a reservoir-mediated entanglement buildup between two qubits.

\paragraph*{Floquet influence functionals}

\begin{figure}
    \centering
    \includegraphics[scale=1]{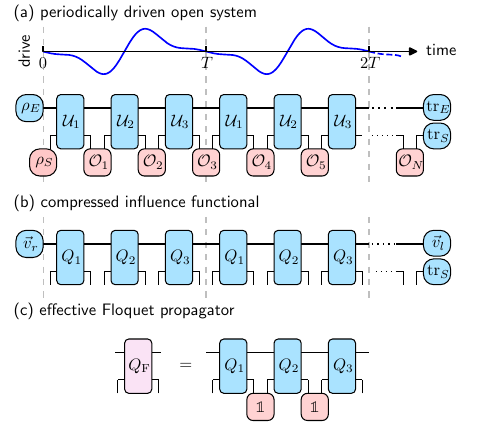}
    \caption{Quantum circuit representation of a periodic Floquet influence functional. (a) Local multi-time correlation functions for open system dynamics with factorized initial conditions. Due to the periodic driving, the global unitaries $\mathcal{U}_n$ are periodically repeating. (b) Periodic compressed MPS representation of the influence functional. The global unitaries are replaced by compressed tensors $Q_n$, but the periodic structure is preserved. (c) Effective time-independent Floquet propagator $Q_\mathrm{F}$, created by contracting all $Q_n$ tensors in one period with identities as system interventions. $Q_\mathrm{F}$ describes the stroboscopic time evolution and defines a stroboscopic Floquet-IF.}
    \label{fig:sketch}
\end{figure}

We will first introduce the general notion of periodic (Floquet) influence functionals, starting from a generic driven quantum dynamics for system and environment with arbitrary Hamiltonian $H(t)$. We assume that the time-dependence is periodic with period $T$, i.e.~$H(t)=H(t+T)$. For convenience we consider the dynamics on a discrete time-grid using time steps $t_n=n\delta t$ chosen such that the driving period is an integer multiple of the time step $T=M\delta t$, with $M\in \mathbb{N}$. Consider the general multi-time correlation function for a set of arbitrary local operators $\mathcal{O}_n$ acting on the system density matrix. Assuming factorized initial conditions, we can write
\begin{equation}\label{eq:cf}
\begin{split}
        \braket{\mathcal{O}_N(N\delta t)&\ldots \mathcal{O}_2(2\delta t) \mathcal{O}_1(\delta t)}=\\&
    \tr\, \mathcal{O}_N\mathcal{U}_N\cdots \mathcal{O}_2 \mathcal{U}_2\mathcal{O}_1\mathcal{U}_1(\rho_\mathrm{sys}\otimes\rho_\mathrm{env})
\end{split}
\end{equation}
where $\mathcal{U}_n$ are global unitary channels generated by the Hamiltonian during the $n$'th evolution step. Choosing a basis for all local operators $\mathcal{O}_n$ we can define the influence functional \cite{feynman63} (IF) $\mathcal{I}$, also known as the influence matrix \cite{Sonner2021Dec} or process tensor \cite{Jorgensen2019Dec,backerVerifyingQuantumMemory2025}, in terms of a large tensor \cite{Pollock2018Jan,Cygorek2021Jan,Sonner2021Dec}
\begin{equation}\label{eq:if_def}
    \mathcal{I}^{(\mu_N\mu_N')...(\mu_1\mu_1')}=\tr\,\mathcal{U}_N^{\mu_N\mu_N'}\cdots\mathcal{U}^{\mu_2\mu_2'}_2\mathcal{U}_1^{\mu_1\mu_1'}\rho_\mathrm{env}.
\end{equation}
The Greek indices $\mu=1\ldots \mathrm{dim}(\mathcal{H}_\mathrm{sys})^2$ label the Liouville-space (density matrix space) of the system, and can be contracted with the operators $\mathcal{O}_n$ to recover the corresponding local observable. Note that, in the IF, the environment degrees of freedom are completely traced out and are no longer accessible. In fact, the right-hand side of \eqref{eq:if_def} provides a special representation of the IF as a matrix product operator (MPO), where the environment degrees of freedom represent a virtual space that expresses the full tensor as a product of matrices, in the sense that for a given index $(\mu,\mu')$, $\mathcal{U}^{\mu\mu'}$ is a matrix. The dimension of this matrix is the so-called MPO bond dimension, in this case corresponding to the full dimension of the environment state space. A key observation here is that, due to the periodicity of the driving, the channels obey $\mathcal{U}_{n+M}=\mathcal{U}_n$, and hence the individual tensors in the IF-MPO, are periodically repeating after $M$ steps \cite{Lerose2021May}, see Fig.~\ref{fig:sketch}a.
Defining a Floquet propagator via $\mathcal{U}_{\mathrm{F}}=\mathcal{U}_M\cdots\mathcal{U}_1$ we can formulate a stroboscopic Floquet-IF \cite{Lerose2021May} as
\begin{equation}
    \mathcal{I}_{\mathrm{F}}^{(\mu_K\mu_K')...(\mu_1\mu_1')}=\tr\,\mathcal{U}_{\mathrm{F}}^{\mu_K\mu_K'}\cdots\mathcal{U}^{\mu_2\mu_2'}_{\mathrm{F}}\mathcal{U}_{\mathrm{F}}^{\mu_1\mu_1'}\rho_\mathrm{env}.
\end{equation}
Within the standard Floquet framework, one usually defines a Floquet Hamiltonian via $\mathcal{U}_{\mathrm{F}}=\exp(-\ii [H_{\mathrm{F}},\cdot] T)$  \cite{bukov2015Universal, Mori2023Mar} generating a dynamical semi-group on the coarse-grained time grid. This general microscopic construction clearly shows that a representation for the IF in terms of a time-periodic MPO exists, but it cannot be used in practice because the environment Hilbert space is typically too large to allow for direct calculations. In the following, we demonstrate how to construct a compressed MPO representation of the IF that retains the periodic structure yet remains computationally tractable through a reduced bond dimension.

\paragraph*{Local driving}
In this work we focus on the important case where the driving acts locally on the system, and provide an outlook for expansion to more general driving protocols later.
For local driving the Hamiltonian takes the form
\begin{equation}\label{eq:Htot}
    H(t)=H_\mathrm{sys}(t)+H_\mathrm{env}+H_\mathrm{int}.
\end{equation}
To be specific, we consider Gaussian bosonic environments represented as a collection of bosonic modes $b_\lambda$ that are coupled linearly to the system via a hermitian coupling operator $S$
\begin{equation}\label{eq;undriven_if}
    H_\mathrm{env}=\sum_\lambda \omega_\lambda b_\lambda^\dagger b_\lambda,\quad 
    H_\mathrm{int}=S\sum_\lambda g_\lambda (b_\lambda^\dagger+ b_\lambda).
\end{equation}
If the system Hamiltonian is absent, the evolution is described by a dynamical semi-group. As described in Ref.~\cite{Link2024May}, this allows to represent the IF of the undriven problem as a matrix product state with repeating tensors
\begin{equation}\label{eq:sgif}
    \mathcal{I}_{\mathrm{undriven}}^{(\mu_N\mu_N')\ldots(\mu_1\mu_1')}= \vec{v}_l\cdot q^{(\mu_N\mu_N')}\cdots q^{(\mu_1\mu_1')}\vec{v}_r,
\end{equation}
where $q^{(\mu_1\mu_1')}$ are square matrices that are independent of $N$ and act on a auxiliary (bond) space, replacing the original full environment state space. The boundary vectors $\vec{v}_{l/r}$ also live in the auxiliary space and resemble the initial state and trace of the environment. 
For Gaussian bosonic baths, this form can be obtained using either tailored Markovian embedding methods such as HEOM \cite{Tanimura2020Jul} or pseudomodes \cite{Mascherpa2020May,huang2025Coupled}, or the uniform variant of the time-evolving matrix product operator method (uniTEMPO) \cite{Link2024May}. Crucially, the required bond dimension representing the number of auxiliary degrees of freedom is small for many realistic environments \cite{Vilkoviskiy2024May,thoenniss2024Efficient}. The uniTEMPO algorithm that we will utilize here can generate a uniform MPO representation as in \eqref{eq:sgif} automatically up to a desired accuracy with typically low bond dimensions (high accuracy leads to larger bond dimension). Previous tensor network algorithms for constructing IFs used finite MPO compression \cite{Banuls2009Jun,Strathearn2018Aug,Jorgensen2019Dec,Cygorek2021Jan,Lerose2021May,Ye2021Jul,Sonner2021Dec,Gribben2022Feb,Thoenniss2023May,Chen2024Jan,Frias-Perez2024Mar,Cygorek2024Apr,Cygorek2024May2} which does not preserve semi-group form and would not permit the following derivations.  

The tensor $q$ generated by the uniTEMPO algorithm describes the evolution in the absence of the system Hamiltonian. To incorporate the local evolution in the IF, we assume a small time-step $\delta t$ and perform a symmetric Trotter factorization of the full unitary evolution.  Given the uniform MPO representation of the IF in Eq.~\eqref{eq:sgif}, we can define a propagator for the $n$th time step as
\begin{equation}
\begin{split}
&({Q}_n)^{\mu\mu'}_{rr'}=\\&\sum_{\alpha\alpha'} \mathcal{U}_\mathrm{sys}^{\mu\alpha}(t_{n},t_{n}-{\delta t}/{2})q^{(\alpha\alpha')}_{rr'}\,\mathcal{U}_\mathrm{sys}^{\alpha'\mu'}(t_{n}-{\delta t}/{2},t_{n-1}),
\end{split}    
\end{equation}
where the indices $r,\,r'$ label the MPO bonds and $\mathcal{U}_\mathrm{sys}(t_f,t_i)$ is the unitary channel generated by $H_\mathrm{sys}(t)$. Due to the effective dynamical semi-group of the environment, this propagator inherits the periodicity of the drive $Q_{n+M}=Q_n$, and hence directly replaces the global unitary $\mathcal{U}_n$ in Eq.~\eqref{eq:if_def}, but as a genuinely dissipative evolution in a much smaller effective state-space. We can likewise define a stroboscopic propagator $Q_\mathrm{F}=Q_M\cdots Q_1$, see Fig.~\ref{fig:sketch}c, which we can directly compute when the environment is efficiently compressed. The eigenvalues of $Q_\mathrm{F}$ define an effective complex Floquet-spectrum, similar to Lindblad-Floquet theory \cite{Schnell2020Mar,Chen2024May}, describing the dynamical time-scales of the system at stroboscopic times. Choosing a different initial phase of the drive does not alter the Floquet spectrum, making it a gauge-invariant property similar to the eigenenergies of $H_\mathrm{F}$ in a unitary setting. The eigendecomposition can be utilized to propagate the system to large evolution times without linear-in-time effort \cite{sonner2025Semi}. For instance, the asymptotic state of the system can be obtained directly from the leading eigenvector of $Q_\mathrm{F}$. Crucially, this allows us to study the asymptotic dynamics of driven dissipative quantum systems directly and beyond linear response. Note, however, that for dissipative dynamics already in the Markovian case it is not possible in general to define a generator analogous to the Floquet Hamiltonian that defines a micro-motion within the physical state space \cite{Schnell2020Mar}.

The numerical convergence of the uniTEMPO algorithm, used to construct the Floquet-IF, is governed by the bond dimension $\chi$ and the Trotter time step $\delta t$. For the following example calculations we have performed convergence analysis with respect to both of these parameters and chosen appropriate values for each considered model such that the results are numerically exact.

\paragraph*{Driven spin-boson model}

\begin{figure}
    \centering
    \includegraphics{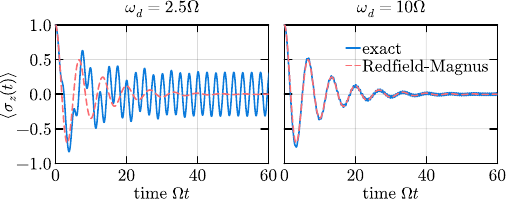}
    \caption{Spin-boson dynamics with transversal driving $H_{\mathrm{drive}}(t) = \epsilon_d\cos(\omega_d t)\sigma_z$ for $\alpha = 0.05$, $\omega_c = 2.5\Omega$, $\epsilon_d = 1\Omega$. We compare the exact uniTEMPO calculation with a Redfield master equation based on the Magnus expansion in the case of slow (left panel) and fast (right panel) driving. As expected, the Magnus expansion provides more accurate result when the driving frequency is high. IF simulations were performed with time-step $\delta t = \pi/(60\Omega)$ and bond dimension $\chi = 235$.}
    \label{fig:quench_dynamics}
\end{figure}

We now show applications of our method to the driven spin-boson model. This paradigmatic model consists of a single qubit driven via a time-periodic field and coupled to a non-interacting bosonic reservior, as in Eq.~\eqref{eq:Htot} \cite{Koski2018Jul,vonHorstig2024Jul,ivakhnenko2025Probing}. 
We choose specifically the coupling operator $S=\sigma_z$ and system Hamiltonian
\begin{equation}
    H_\mathrm{sys}(t)=\frac{\Omega}{2}\sigma_x+H_\mathrm{drive}(t).
\end{equation}
We will consider two different types of external driving, longitudinal driving $H_\mathrm{drive}(t) = \epsilon_d\cos(\omega_d t)\sigma_x$ and transversal driving $H_\mathrm{drive}(t) = \epsilon_d\cos(\omega_dt)\sigma_z$, with driving amplitude $\epsilon_d$ and driving frequency $\omega_d$. We assume throughout an Ohmic bath at zero temperature with exponential high-frequency cutoff 
\begin{equation} \label{sd}
    J(\omega) = \sum_\lambda |g_\lambda|^2\delta(\omega-\omega_\lambda)=\alpha\omega\mathrm{e}^{-\omega/\omega_c},
\end{equation}
where $\alpha$ is a dimensionless parameter that determines the system-bath coupling strength and $\omega_c$ is a cutoff frequency. When the driving is slow or the coupling to the bath is strong, simple quantum master equation approaches fail to provide accurate predictions for the reduced dynamics of this model. As an example we consider a master equation approach that follows from performing the Floquet-Magnus expansion of the full Hamiltonian \cite{casas2001Floquet,rahav2003Effective,bukov2015Universal,dey2025Error} and then deriving a standard Redfield equation of this effective system-bath model \cite{Hartmann2020}. In the first order of the Magnus-expansion one simply obtains a modified Hamiltonian describing an undriven spin-boson model with renormalized parameters. The periodic micromotion can then be recovered via suitable kick-operators. In Fig.~\ref{fig:quench_dynamics} we display exact quench dynamics of a driven spin-boson model obtained from a Floquet-IF generated via uniTEMPO and compare it to the solution of the Redfield-Magnus approach. For the considered parameters, the master equation accurately describes the dynamics of $\braket{\sigma_z}$ for high-frequency driving. However, it fails in the case of slower driving due to a failure of the underlying Magnus expansion, suggesting that the system can no longer be described by an effective equilibrium problem, i.e.~there may no longer be an effective Floquet-Gibbs state that the system relaxes towards \cite{Shirai2016May, Mori2023Mar}. Note that more restrictive approaches, such as Floquet-Lindblad master equations \cite{Lendi1986May, Breuer2009Nov, Schnell2020Mar, kolisnyk2024Floquet} perform very badly for this example, with a Markovian description of the reduced dynamics becoming even more restrictive in driven systems. We leave a benchmarking of further master equation approaches \cite{Mori2023Mar} against our exact numerics to future works. 

\paragraph*{Floquet heating}

\begin{figure}
    \centering
    \includegraphics[scale = 1]{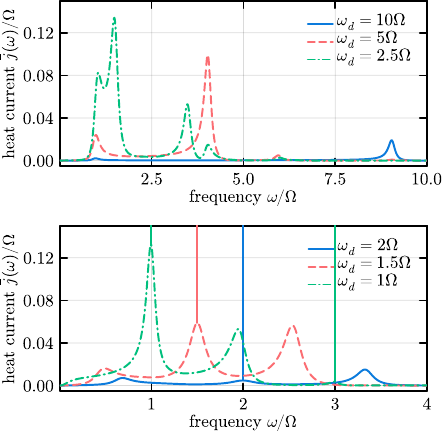}
    \caption{Time-averaged asymptotic heat current density in a strongly driven spin-boson model ($\alpha = 0.05$, $\omega_c = 2.5\Omega$, $\epsilon_d = 1\Omega$) for longitudinal $H_{\mathrm{drive}}(t) = \epsilon_d\cos(\omega_d t)\sigma_x$ (top panel) and transversal $H_{\mathrm{drive}}(t) = \epsilon_d\cos(\omega_d t)\sigma_z$ (bottom panel) driving fields with different driving frequencies. In the transversal driving case, resonant excitations at odd multiples of the driving frequency lead to delta peaks, indicated by straight lines. Simulations were performed with time-step $\delta t \approx\pi/(60\Omega)$ and bond dimension $\chi = 235$.}
    \label{fig:heat_current}
\end{figure}

To further characterize the non-equilibrium properties of the driven spin-boson model we study Floquet-heating of the bath. Periodic driving typically pumps energy to the system, competing with the dissipation induced by the environment \cite{kato2016Quantum,Gribben2022Oct, albarelli2024Pseudomode}. In thermal equilibrium the energy flowing from the system towards any bath mode vanishes exactly, making the heating of the bath a strong signature of departure from equilibrium. Moreover, a detailed characterization of Floquet heating is useful for the purpose of dissipative Floquet-engineering, where an efficient cooling of the system due to the reservoir is necessary for stabilizing long-time dynamics which is otherwise limited by thermalization towards a trivial infinite temperature state \cite{eckardt2017Colloquium, Ritter2025}.
The heating of bath modes of a given frequency can be characterized by a frequency-resolved heat current density, given by the change in the corresponding mode energy 
\begin{equation}
    j(t,\omega) = \sum_\lambda \frac{\diff}{\diff t}\langle\omega_\lambda b^\dagger_\lambda(t) b_\lambda(t)\rangle \delta(\omega-\omega_\lambda).
\end{equation}
For Gaussian baths this function can be expressed in terms of local two-point correlation functions \cite{Gribben2022Oct, albarelli2024Pseudomode}, which we can efficiently determine for arbitrary evolution times with our influence functional approach. We stress that obtaining accurate correlation functions is challenging using master equations, already in absence of any driving \cite{koyanagi2024Classical, keeling2025Process}. We consider here specifically the heat current density averaged over one driving period
\begin{equation}\label{eq:av_heat_current}
    \bar j(\omega)=\frac{1}{T}\int_{t}^{t+T} dt' j(t',\omega).
\end{equation}
In the quasi-stationary regime $t\rightarrow\infty$ this averaged current density is independent of the observation time. In our simulations we first determine the stationary state from the leading eigenvector of the Floquet propagator $Q_\mathrm{F}$ and then evolve the relevant correlation functions with the help of the micro-motion propagators $Q_n$. Further details on computing this quantity are provided in the End Matter. We display examples for heat current densities in Fig.~\ref{fig:heat_current}. The upper panel shows heat currents in the longitudinal driving case for different driving frequencies. As expected, for larger driving frequencies the heat currents decreases as the system is closer to an effective equilibrium. Transport is amplified at resonant frequencies $n\omega_d\pm \Omega$. For transversal driving, the heat current density exhibits features reminiscent of resonance fluorescence, where coherent scattering yields a delta peak at the driving frequency, while inelastic processes produce a Mollow-triplet structure with sidebands shifted by the Rabi frequency \cite{wallsBook}, here equal to the driving amplitude $\epsilon_d$. Our exact calculations likewise show delta peaks at odd multiples of $\omega_d$, corresponding to resonant energy transport with finite total heat (computed exactly, see End Matter). Around each multiple $n\omega_d$, most prominently for $n=1$, we find a broadened central peak and sidebands near $\omega_d \pm \epsilon_d$, qualitatively reflecting the Mollow triplet.

\begin{figure}
    \centering
    \includegraphics[scale = 1]{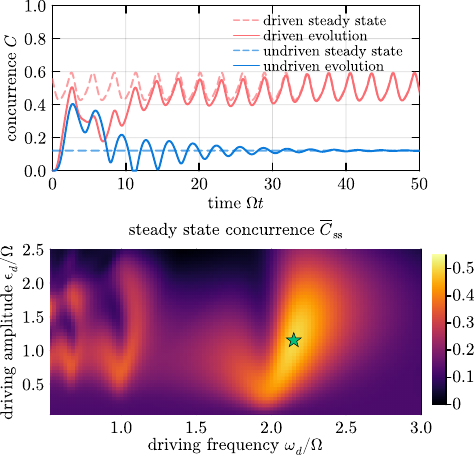}
    \caption{Entanglement dynamics in a driven two-spin-boson model Eq.~\eqref{eq:two_spin_H} with $\alpha = 0.1$, $\omega_c = 5\Omega$. Top panel: Entanglement dynamics after a quench for the initial state $\rho_0 = \ket{00}\!\bra{00}$ in the undriven ($\epsilon_d = 0$) and driven ($\omega_d = 2.15\Omega$, $\epsilon_d = 1.15\Omega$, star in lower panel) cases.
    Bottom panel: period-averaged concurrence of the Floquet steady state as a function of the driving frequency and the driving amplitude. Simulations were performed with time-step $\delta t \approx \pi/(48\Omega)$ and bond dimension $\chi = 342$.}
    \label{fig:concurrence}
\end{figure}

\paragraph*{Driving-assisted entanglement stabilization}
As a second application, we investigate environment-mediated entanglement generation in a driven two-spin-boson model. We consider two uncoupled qubits ${A}$ and ${B}$ interacting with a common bosonic environment through the coupling operator $S=\frac{1}{2} (\sigma_z^A + \sigma_z^B)$. Additionally, we include a local transversal driving to both qubits
\begin{equation} \label{eq:two_spin_H}
    H_\mathrm{sys}(t) = \frac{\Omega}{2} (\sigma_x^A + \sigma_x^B) +\frac{\epsilon_d}{2}\cos(\omega_d t) (\sigma_x^A + \sigma_x^B).
\end{equation}
We assume initial conditions to lie in the (symmetric) triplet-subspace only. We consider once again the Ohmic bath defined via Eq.~\eqref{sd}. When the system is not driven ($\epsilon_d=0$) and temperature is low the qubits become entangled through the interaction with a common environment \cite{Hartmann2020Oct}. An example of this entanglement buildup after a quench is displayed in Fig.~\ref{fig:concurrence} (upper panel, blue line). After quenching from a separable initial state, the concurrence of the two qubits increases to around 0.4 and then approaches the equilibrium value of around 0.1. Our goal now is to use the local driving in \eqref{eq:two_spin_H} in order to realize larger values for the asymptotic entanglement \cite{Patrick2010, Schwartz2016}. As can be seen in Fig.~\ref{fig:concurrence}, with optimally chosen driving parameters we are able to reach values of around 0.5 for the asymptotic concurrence. In order to find these optimal parameters we computed the concurrence of the asymptotic Floquet steady state as a function of the driving amplitude $\epsilon_d$ and driving frequency $\omega_d$, displayed as a heat-map in the lower panel of Fig.~\ref{fig:concurrence}. We find significant enhancement of the qubit entanglement at around $\omega_d\approx2\Omega$ and moderate driving amplitudes. This frequency closely resembles the oscillations of the concurrence observed at short times after the quench in the undriven dynamics, suggesting that the driving stabilizes the transient entanglement buildup via resonantly driving a transition to a strongly entangled short-lived state. In fact, as we explain in the End Matter, at weaker driving the optimal driving frequency can be estimated directly form a spectral analysis \cite{Link2024May, sonner2025Semi} of the transfer matrix $q$ in Eq.~\eqref{eq;undriven_if}, avoiding the full parameter scan in Fig.~\ref{fig:concurrence}.

\paragraph*{Conclusions} The Floquet-IF framework can be used to describe the dynamics of driven open systems in both intermediate and strongly damped regimes. As we have shown, it can be applied to identify optimal driving protocols that steer a system into otherwise short-lived states with desirable properties. 
One limitation of our numerical technique is that we require the driving to be local (i.e. acting only on the system part) in order to construct a Floquet-IF from a semi-group IF. This excludes the important case of time-dependent coupling operator, often encountered in thermodynamical cycles \cite{Quan2007, Kosloff2017,koyanagi2022Numerically,boettcher2024Dynamics,alamoMinimalQuantumHeat2024}. We propose a way to overcome this issue via the reaction coordinate (RC) mapping, which we further explain in Sec.\ref{rc_mapping} of the End Matter. 
In the future, the method could serve as a numerical tool to study the dissipative stabilization of Floquet-engineered states over long evolution times \cite{Ritter2025}. Access to multi-time statistics makes it possible to characterize non-equilibrium features in driven dissipative systems in detail, which is especially relevant for studies of transport and quantum thermodynamics \cite{binderThermodynamicsQuantumRegime2018}. Crucially, influence functionals can be directly included in many-body systems with local reservoirs, making extensions to larger interacting systems possible \cite{Fux2022Jan,keeling2025Process}. Moreover, the presented IF-MPO approach could be used as a fast and accurate real-time impurity solver \cite{nayakSteadystateDynamicalMean2025a} for dynamical mean-field theory (DMFT) studies of driven strongly correlated systems \cite{tsujiCorrelatedElectronSystems2008,aokiNonequilibriumDynamicalMeanfield2014,scarlatellaDynamicalMeanFieldTheory2021}. 
\paragraph*{Acknowledgments} VL acknowledges supported by the Deutsche Forschungsgemeinschaft (DFG, German Research Foundation) via the Research Unit FOR 5688 (Project No. 521530974).

\paragraph*{Data availability} The data that support the findings of this article are openly available at \cite{mickiewicz2026data}.

%

\clearpage
\section*{End Matter} \label{end_matter}

\subsection{Influence functional formalism}

We provide details on the influence functional  formulation of open system dynamics \cite{keeling2025Process}. We define the quantum channel that propagates a general state of system and environment from time $t_{n-1}$ to $t_{n}$ via
\begin{equation}\label{eq:channel}
    \mathcal{U}_n\rho=U(t_{n},t_{n-1})\rho U^\dagger(t_{n},t_{n-1})
\end{equation}
where $U(t_f,t_i)$ are the standard unitary operators generated by $H(t)$. Formally, one can define the influence functional $\mathcal{I}$ as the multi-linear map from a set of local superoperators $\{\mathcal{O}_n\}$ and the system initial state $\rho_\mathrm{sys}$ to the corresponding multi-time correlation function \eqref{eq:cf}
\begin{equation}
    \mathcal{I}[\{\mathcal{O}_n\},\rho_\mathrm{sys}]=\braket{\mathcal{O}_N(N\delta t)\ldots \mathcal{O}_2(2\delta t) \mathcal{O}_1(\delta t)}.
\end{equation}
In order to obtain lower-order multi-time observables, one chooses $\mathcal{O}_m=\id$ for a subset of superoperators. For instance, time-local observables as encoded in the reduced system state $\rho_\mathrm{sys}(t)$ can be obtained from setting all but one operator to the identity.
\begin{equation}
    \tr_\mathrm{sys} \mathcal{O}_n\rho_\mathrm{sys}(n\delta t)=\mathcal{I}[\id,\ldots ,\id,\mathcal{O}_n,\id\ldots,\id;\rho_\mathrm{sys} ].
\end{equation}
We use throughout the superoperator formulation of quantum dynamics, where operators are mapped to vectors. Specifically, we choose a basis for the system Hilbert space such that local operators (e.g.~the reduced density matrix) are vectorized $\rho_\mathrm{sys}^\mu=\braket{\mu^+|\rho_\mathrm{sys}|\mu^-}$, with $\mu\equiv(\mu^+,\mu^-)$ taking values from $1,\ldots ,\mathrm{dim}(\mathcal{H}_\mathrm{sys})^2$.
Choosing operators $\mathcal{O}_n$ to be individual basis elements, i.e.~$\mathcal{O}_n\rho_\mathrm{sys}:=\ketbra{(\mu_n')^{+} }{\mu_{n-1}^{+}}\rho_\mathrm{sys}\ketbra{\mu_{n-1}^{-} }{(\mu_n')^{-}}$, leads to the tensor representation \eqref{eq:if_def}. The path-conditioned operators $\mathcal{U}_n^{\mu\nu}$ are superoperators in the environment Hilbert space and are given explicitly as
\begin{equation}
    \begin{split}
&\mathcal{U}^{\mu_n\mu_n'}_n\rho_\mathrm{env}=\\&\braket{\mu_n^+|U(t_{n},t_{n-1})|(\mu_n')^+}\rho_\mathrm{env}\braket{(\mu_n')^-|U^\dagger(t_{n},t_{n-1})|\mu^-_n}
\end{split}
\end{equation}
according to Eq.~\eqref{eq:channel}.

\subsection{Heat current density}

For bosonic environments, the heat current density at time $t$ is given as \cite{Gribben2022Oct, albarelli2024Pseudomode}
\begin{equation}\label{eq:heat_current}
\begin{split}
    j(t, \omega) = 2J(\omega)\omega \int_{0}^t ds& \mathrm{Im} \{ ((1+2n_B(\omega))\sin(\omega (t-s))\\
                   & + \ii \cos(\omega (t-s))) \langle S(t)S(s)\rangle\}.
\end{split}
\end{equation}
On order to remove the dependence on the final time $t$ we consider the period-averaged heat current \eqref{eq:av_heat_current}. At asymptotic times $t\rightarrow\infty$. 
one obtains the following expression for the average heat current
\begin{equation}\label{eq:av_heat_current_expl}
\begin{split}
    \overline{j}(\omega)={2J(\omega)\omega} \int_0^\infty d\tau \mathrm{Im} \{ (&(1+2n_B(\omega))\sin(\omega \tau) \\
    +& i \cos(\omega \tau)) \overline{C}(\tau)\}.
\end{split}
\end{equation}
where the period-averaged asymptotic correlation function is
\begin{equation}
    \overline{C}(\tau)  =\frac{1}{T} \int_{t}^{t+T} dt'\langle S(t'+\tau)S(t')\rangle.
\end{equation} 
The primary computational challenge lies in the evaluation of this correlation function in the Floquet steady state, which can be achieved efficiently with the methods introduced in the main text. 

For systems with a finite memory time the correlation function will factorize for large $\tau$
\begin{equation}
     \langle S(t+\tau)S(t)\rangle \rightarrow \langle S(t+\tau)\rangle \langle S(t)\rangle.
\end{equation}
Note that the expression on the right hand side may or may not be zero, depending on the details of the model. In order to determine the integral in Eq.~\eqref{eq:av_heat_current_expl} we split the correlation function into a connected and factorized term
\begin{equation}
    \overline{C}(\tau) = \overline{C}_\mathrm{decay}(\tau) + \overline{C}_\mathrm{asym}(\tau),
\end{equation}
where 
\begin{equation}
    \overline{C}_\mathrm{asym}(\tau)=\frac{1}{T}\int_t^{t+T}\diff t' \langle S(t+\tau)\rangle \langle S(t)\rangle.
\end{equation}
The connected correlation function $\overline{C}_\mathrm{decay}$ decays for large $\tau$, allowing for a direct numerical integration. The remaining term $\overline C_\mathrm{asym}$ is $T$-periodic and can therefore be expanded as a Fourier series. In fact, since $\braket{S(t)}$ is periodic and real the expansion takes the form
\begin{equation} \label{fourier}
    \overline C_{\mathrm{asym}}(\tau) = \sum_{n=0}^\infty c_n\cos(n \omega_d \tau),
\end{equation}
with positive, real coefficients $c_n$.
Inserting this in Eq.~\eqref{eq:av_heat_current_expl} one finds delta-peak contributions to the heat current density
\begin{equation} \label{asym_hc}
\overline{j}_{\mathrm{asym}}(\omega) = \pi J(\omega)\omega \sum_{n=0}^\infty c_n  \delta(\omega - \omega_dn).
\end{equation}

\subsection{Total energy transfer}
\begin{figure}
    \centering
    \includegraphics{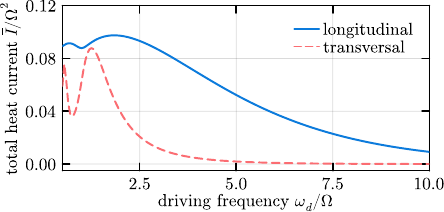}
    \caption{Total time-averaged heat current \eqref{eq:curr_tot} as a function of the driving frequency with the same parameters as in Fig.~\ref{fig:heat_current}.}
    \label{fig:total_hc}
\end{figure}

The total amount of heat transferred from the system to the bath is given by the total heat current
\begin{equation}\label{eq:curr_tot}
    \overline{I}= \int_0^\infty d\omega \;\overline{j}(\omega).
\end{equation}
In Fig. \ref{fig:total_hc} we display the total heat current as a function of the driving frequency. While at low driving frequencies, resonances with the system transitions lead to a non-monotonic behavior, the heat current follows a smooth asymptotic towards zero as the driving frequency is increased. At large frequencies the driving no longer influences the dynamics due to the large detuning from all resonances. Then the system is described by the undriven equilibrium state with vanishing currents. However, in the case of longitudinal driving, the heat current vanishes significantly slower compared to transversal driving. 

\subsection{Spectral analysis of the transient entanglement in a two-spin-boson model}

In this section we provide an additional analysis of the entanglement stabilization in the locally driven two-spin-boson model. We show that this effect can be understood in terms of a resonant driving of a transient state that can be identified in the spectrum of the full transfer matrix. For the undriven dynamics, the transfer matrix ($q$ in Eq.~\eqref{eq;undriven_if}) represents a dissipative propagator with a spectrum similar to that of a Lindbladian, but now in a non-Markovian system where the number of dynamical time-scales is not restricted by the local system dimension. The reduced state after a quench can be decomposed as \cite{Link2024May}
\begin{equation}
    \rho(t) = \sum_{n=1}^{\chi d^2} \rho_n \mathrm{e}^{\gamma_nt} 
\end{equation}
where complex rates $\gamma_n$ and states $\rho_n$ are defined through eigenvalues and eigenstates of the propagator $q$. The contribution $\rho_1$ with a zero eigenvalue $\gamma_1 = 0$ is the steady state with $\tr\rho_1 = 1$, whereas other terms obey $\tr\rho_{n\neq 1}=0$ and $\mathrm{Re}\gamma_{n\neq 1} < 0$, and hence describe transient dynamics. We can estimate the contribution of a single eigenstate to the system state via $\rho^{(m)}(t) = \rho_1+\rho_m\mathrm{e}^{\ii\mathrm{Im}\gamma_m t} + \rho^\dagger_m\mathrm{e}^{-\ii\mathrm{Im}\gamma_m t}$, where we include the steady state to ensure $\rho^{(m)}$ is normalized.

\begin{figure} [H]
    \centering
    \includegraphics[scale = 1]{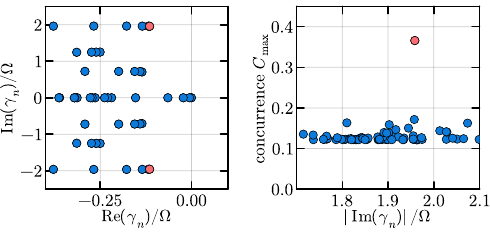}
    \caption{Spectral analysis of undriven entanglement dynamics with the same parameters as in Fig.~\ref{fig:concurrence}. We compute a spectrum of the undriven propagator (left panel) and the maximal contribution of single eigenvectors to concurrence (right panel). The eigenstate marked in red has a particularly large contribution to concurrence.}
    \label{fig:spectrum}
\end{figure}

In the left panel of Fig.~\ref{fig:spectrum} we display the dominant spectral region for the transient dynamics. In the right panel of Fig.~\ref{fig:spectrum} the maximum concurrence of the corresponding states $\rho^{(m)}$ is displayed. We find a single eigenvector with large concurrence values marked in red, at a frequency $\mathrm{Im}\gamma\approx 2\Omega$. 
Thus, we can account the transient entanglement buildup during the quench displayed in Fig.~\ref{fig:concurrence} to this particular spectral contribution. The
entanglement is stabilized in the driven system via resonant driving $\omega_d\approx \mathrm{Im}\gamma$ of a transition to this state. In fact, invoking this spectral analysis would enable us to find a good guess for a suitable driving frequency without the expensive full parameter scan from Fig.~\ref{fig:concurrence}.

\subsection{Periodic coupling via the reaction coordinate mapping} \label{rc_mapping}

The approach presented above assumes that the time-dependence is in the system Hamiltonian $H_\mathrm{sys}$ only. Here we describe a method to construct a Floquet-IF for cases where instead the coupling is time-dependent, by using the reaction coordinate (RC) mapping \cite{Hughes2009Jul,chin2010Exact,iles2014Environmental,strasberg2018Fermionic,restrepo2018From,nazir2019The,shubrook2025Non}. In the RC mapping the bath modes are transformed linearly to a new set of modes, consisting of the reaction coordinate $a_0$ and the residual bath ($a_\lambda$ modes). The transformation is designed in such a way that the system couples only to the RC mode
\begin{equation}
    H_\mathrm{int}(t)=S(t)\sum_\lambda g_\lambda (b_\lambda+b_\lambda^\dagger)=S(t)g_0(a_0+a_0^\dagger).
\end{equation}
The bath Hamiltonian then includes a hopping between the RC mode and the residual bath modes
\begin{equation}
    H_\mathrm{env}=\omega_0 a_0^\dagger a_0+\sum_\lambda \tilde g_\lambda(a_0 a_\lambda^\dagger+ a_0^\dagger a_\lambda ) + \sum_\lambda \tilde\omega_\lambda a_\lambda^\dagger a_\lambda. 
\end{equation}
This allows us to include the RC mode into the system, while treating the residual bath as another stationary environment for which a semi-group IF can be constructed. The reaction coordinate can then be absorbed back into the bath IF, similar to the procedure in Ref.~\cite{Cygorek2021Jan}, recovering a periodic Floquet-IF for the original system.
\end{document}